# Unified Quantum Classical Theory of Einstein Diffusion-Mobility Relationship for Ordered and Disordered Semiconductors

K. Navamani and Swapan K. Pati
*Theoretical Sciences Unit*
*Jawaharlal Nehru Centre for Advanced Scientific Research*
*Jakkur-560064, Bangalore, INDIA*

We propose a unified diffusion-mobility relation which quantifies both quantum and classical levels of understanding on electron dynamics in ordered and disordered materials. This attempt overcomes the inability of classical Einstein relation (diffusion-mobility ratio) to explain the quantum behaviors, conceptually well-settles the dimensional effect, phase transition and nonlinear behavior of electronic transport. Our proposed theory relies on the chemical potential which provides the coupling mechanism of charge-heat current, due to electron-phonon coupling. We have derived expressions which explain charge transport in both degenerate and nondegenerate materials, and also provide the linear and nonlinear relationship between the charge density and chemical potential. Theoretically, we find that the symmetrical nature of electron-hole transport in strongly correlated two-dimensional semiconductors indicates linear dispersion. We also observe the broken symmetry in the nonlinear regime. This generalized diffusion-mobility relation explains both the strongly and weakly correlated systems from low temperature to high temperature, in both the relativistic as well as nonrelativistic domains. In vanishing charge density limit of nondegenerate cases, the nonlinear transport reduces to linear like transport, which is the classical Einstein relation.

The charge transport of semiconductors in thermal equilibrium is given by the classical Einstein relation [1, 2]

$$\frac{D}{\mu} = \frac{k_B T}{q}, \qquad (1)$$

where $D$ is the diffusion coefficient, $\mu$ is the mobility, $k_B$ is the Boltzmann constant, $T$ is the temperature and $q$ is the charge of the particle. The above Einstein relation is valid for non-degenerate, low density semiconducting materials, with no drift current. [3, 4] However, for better device applications, materials with high-charge density are used. [3, 4] Such high charge density leads to strong correlations with collective behavior, thereby showing diffusive electrical transport (Dirac fluid (DF) like behavior). [5-8] In fact, numerous theoretical as well as experimental findings suggest that there is deviation of $D/\mu$ from the classical value of $k_BT/q$, because response of the electric field to diffusion dominate over its response to mobility. [3, 6, 9-12]

A number of reports manifest the quantum nature in electronic materials: phase transition, [13-16] quantum Hall effect, [14, 17-19] filling factor dependent electronic transport, [17, 18, 20] quantum capacitance, [17, 18] particle compressibility, [18, 21, 22] thermoelectricity, [5, 7, 23, 24] etc., to name a few. [16, 21, 25-27] Also, various observations show the nonlinear transport at finite temperature, which indicate the failure of Einstein relation. [6, 8, 19, 25, 28-30] The deviation of diffusion-mobility ratio from classical value was pointed out by Tessler and collaborators; they argued that the diffusion-mobility relation depends not only on the charge density but also on the gradient of charge density and energy transport. [4] Also, they found that the Einstein relation does not fully describes the diode ideality factor. [4, 31] In principle, the various quantum behavior in materials, as mentioned here, can be tuned by the chemical potential with the aid of chemical and electrochemical doping, applying gate voltage, temperature, etc.

Recently, Crossno *et al.* [5] experimentally observed few interesting phenomena in graphene; a) particles in excited states are strongly correlated and show hydrodynamic-like DF behavior, b) very fast electron-electron collisions leading to diffusion enhanced mobility, c) time-reversal invariance assures electron-hole symmetry with linear energy dispersion relation. d) the motion of charge carriers is very fast and is in the relativistic domain which reveals the 2 + 1 dimensional electronic transport, d) the Wiedemann-Franz (WF) law is deviated in both the high temperature as well as in the charge neutrality point in the DF regimes and e) the disorder and temperature strongly affect the electrical transport which limits the diffusion-mobility ratio. But for quasi 2D materials, nonlinear dispersion takes over the electron-hole symmetry, leads to broken symmetry. [19, 32]

On the other hand, the disordered charge transport is observed in organic semiconductors and is explained by the thermally activated hopping process. [33, 34] Here, the presence of disorder breaks down the equilibrium which leads to the deviation from classical Einstein relation. [3, 4, 8, 9, 12, 28] For disordered semiconductors (degenerate materials), the charge equilibrium time is long enough, rendering them in non-equilibrium regime. [35] Recent theoretical investigations on charge transfer kinetics, with the aid of Monte-Carlo (MC) simulations, in disordered organic molecular systems reveal that the charge equilibrium time depends on the amount of disorder and can be characterized by disorder drift time. [34, 36, 37] In these situations, the classical diffusion-mobility relation is limited and it deviates from its actual classical value, $(D/\mu) > (k_B T/q)$.

In this letter, we propose to explain all the above experimental findings with a unified theory. The theoretical framework is based on quantum and classical descriptions of Einstein diffusion-mobility relation using Fermi-Dirac (F-D) distribution function for strongly correlated systems and Maxwellian form for weakly correlated systems, in both the relativistic and nonrelativistic limits. The generalized Einstein relation can be expressed for a general charge-carrier energy-distribution function, and for a general density of states (DOS) function as [2, 3]

$$\frac{D}{\mu} = \frac{n}{q\frac{\partial n}{\partial \eta}}, \quad (2)$$

where, $n$ is the number density (charge carrier density, $\rho = ne$), and $\eta$ is the chemical potential. $n$ gives the particle concentration in the finite region of space. Here, we have formulated the diffusion-mobility relation for 1D, 2D and 3D systems using general DOS functions and F-D distribution functions. Also, we have developed diffusion-mobility formula in classical limits (for low density systems) and compared with high density Fermionic systems. The general expression for the chemical potential dependent number of particles can be written as [3]

$$N(\xi) = \int_{-\infty}^{\infty} f(\varepsilon) g(\varepsilon, \xi) d\varepsilon, \quad (3)$$

where $\varepsilon$ is the normalized energy $\varepsilon = E/k_B T$, $\xi$ is the normalized chemical potential $\eta/k_B T$, $f(\varepsilon)$ is the Fermi-Dirac (F-D) distribution function, and $g(\varepsilon, \xi)$ is the DOS function. The charge carrier density for non-relativistic particle (or Schrödinger particle) in the d-dimensional disordered semiconducting materials can be expressed as,

$$n_d = \left(\frac{1}{\hbar}\right)^d \left(\frac{2mk_B T}{\pi}\right)^{d/2} \sum_{k=1}^{n} (-1)^{k+1} \left(\frac{1}{k}\right)^{d/2} \exp\left(\frac{k\eta}{k_B T}\right) \quad (4)$$

where m is the effective mass of the charge carrier.

By substituting the eq. (4) in eq. (2), one can obtain the general form of diffusion-mobility ratio for 1D, 2D and 3D materials

$$\left.\frac{D}{\mu}\right|_d = \frac{k_B T}{q} \left[ \frac{\sum_{k=1}^{n}(-1)^{k+1}(k)^{-d/2}\exp\left(\frac{k\eta}{k_B T}\right)}{\sum_{k=1}^{n}(-1)^{k+1}(k)^{1-(d/2)}\exp\left(\frac{k\eta}{k_B T}\right)} \right]. \quad (5)$$

For 2D semiconductors, since DOS is energy independent, we derived the charge carrier density analytically without making any approximation, and hence eq. (4) becomes

$$n_{d=2} = \frac{2mk_B T}{\pi \hbar^2} \ln\left(1+\exp\left(\frac{\eta}{k_B T}\right)\right). \quad (6)$$

Thus, the deterministic form of diffusion-mobility relation for 2D disordered semiconductors can be written as

$$\left.\frac{D}{\mu}\right|_{d=2} = \frac{k_B T}{q} \left[ \frac{\left(1+\exp\left(\frac{\eta}{k_B T}\right)\right)\ln\left(1+\exp\left(\frac{\eta}{k_B T}\right)\right)}{\exp\left(\frac{\eta}{k_B T}\right)} \right]. \quad (7)$$

It is to be noted that, at very low temperature and degenerate limit, the normalized chemical potential ($\xi$) is too large, i.e., $\exp(\eta/k_B T) \gg 1$, [7] and hence the charge density of 2D semiconductors can be expressed as $n_{d=2} = 2m\eta/\pi\hbar^2$, accordingly the diffusion-mobility Equation (7) reduces to,

$$\left.\frac{D}{\mu}\right|_{d=2} \approx \frac{\eta}{q}. \quad (8)$$

It has been found that the linear relationship between $n_{d=2}$ and $\eta$ for degenerate systems indicate a linear dispersion. Importantly, eq. 8 is also valid for strongly correlated systems with electron-hole symmetry, which preserve time-reversal invariance (i.e., for electron $\eta \to +\eta$ and for hole $\eta \to -\eta$ in eq. (8)). Here, the diffusion-mobility ratio is purely quantum electronic in nature and depends linearly only on chemical potential, but not on thermal energy. In other words, at the absolute zero of temperature, the diffusion-mobility relation depends linearly on Fermi energy. With increase in temperature, this relation would move towards nonlinear regime, where the chemical potential as well as the thermal energy dictate the diffusion-mobility ratio (eq. (7)). Accordingly, nonlinear relationship between $n_{d=2}$ and $\eta$ causes the broken symmetry (eq. (6)), and it in turn gives rise to nonlinearity in $D/\mu$, which agrees well with the earlier observations. [6, 9-12, 19] The main point is that a system where electron-electron and electron-phonon interactions dominate, the basic energy which determines its electronic properties is the chemical potential. At high temperature, the carrier motion becomes random and the transport is estimated by the thermal averaged diffusion process. In this limit, the classical Einstein relation would remain valid for systems of non-interacting particles. For such weakly correlated (high temperature) situation, the charge transport is weakly dependent on electronic energy, but strongly dependent on thermal energy. Using Eq. (7), we have plotted the $D/\mu$ ratio as a function of $\eta$ for 2D materials at different temperatures which is shown in Fig. 1. We note that the $D/\mu$ sharply increases with $\eta$ in the low temperature regime, while in the high temperature regime it has weak dependence on $\eta$ (nondegenerate cases), reduces to classical Einstein relation. The inset plot shows the quantum to classical transition with respect to $\eta$ while the temperature varies from low to high.

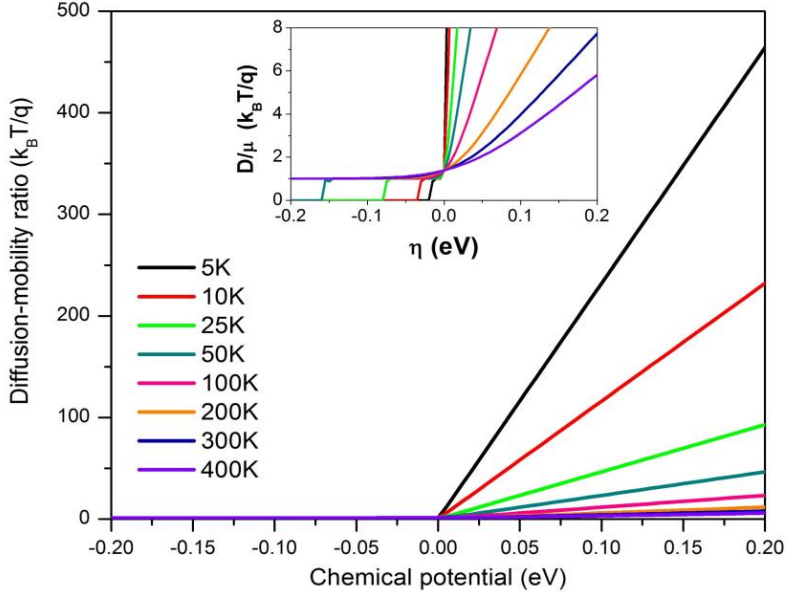

FIG. 1. The $D/\mu$ ratio as a function of $\eta$ at different temperatures. At high temperature, the classical Einstein relation is shown to be valid. The inset plot dictates the classical to quantum transition in the degenerate cases, $\eta \gg k_B T$.

In case of 1D and 3D semiconductors, we used Stirling approximation in eq. (4) to obtain the expressions for charge density and the resulting diffusion-mobility expressions can be written as,

$$\left.\frac{D}{\mu}\right|_{d=1,3} \approx \frac{k_B T}{q}\left[1+\exp\left(\frac{\eta_d}{k_B T}\right)\right]. \qquad (9)$$

As can be seen from eq. (9), the diffusion-mobility ratio for 1D and 3D semiconductors are same except for the dimension dependent chemical potential. For very low density materials (vanishing density limit), the chemical potential satisfies the condition of $\eta_d \to -\infty$, which is more appropriate for the nondegenerate semiconductors. [28] In this limit, the eq. (9) reduces to the classical Einstein relation $(D/\mu)_{1,2,3} = (k_B T/q)$, which is in agreement with earlier experimental work by Wetzelaer *et al.* [35] Further, one can express the chemical potential in terms of dimension dependent carrier density ($n_d$) and thermal wavelength $\left(\lambda = \left(2\pi\hbar^2/mk_B T\right)^{1/2}\right)$ as [38]

$$\eta_d = k_B T \log n_d \lambda^d, \qquad (10)$$

where $d$ is the dimensionality. Inserting eq. (10) in eq. (9), we can modify the generalized diffusion-mobility expression as

$$\left.\frac{D}{\mu}\right|_d \approx \frac{k_B T}{q}\left[1+n_d \lambda^d\right], \qquad (11)$$

Note that, the carrier density ($n_d$) is volume normalized $\left(n_d = N_d/L^d\right)$. At low temperature, the thermal wavelength is large and inter-particle separation is comparable or lesser than the thermal wavelength, and activated charge density is also high, i.e., $n_d \lambda^d \gg 1$. Thus, the eq. (11) can be reduced at low temperature as,

$$\left.\frac{D}{\mu}\right|_d \approx \frac{k_B T}{q} n_d \lambda^d. \qquad (12)$$

In high temperature and non-degenerate regime, $\lambda$ and $n$ are very small (here inter-particle separation is larger than thermal wavelength), and hence $n_d \lambda^d \to 0$. In this context, the diffusion-mobility expression (11) can be reduced to the classical Einstein relation, $(D/\mu)_d \cong (k_B T/q)$, for all dimensional systems.

Using eq. (12) and the thermal wavelength, one can obtain the general expression for "conductivity ($\sigma = nq\mu$)-diffusion (*D*) ratio" at low temperature as,

$$\left.\frac{\sigma}{D}\right|_d = \left[\frac{q^2}{(k_BT)^{1-(d/2)}}\right]\left(\frac{m}{2\pi\hbar^2}\right)^{d/2} \quad (13)$$

It has to be noted from eq. (13) that the conductivity-diffusion relation for 2D materials depends only on the effective mass of the carrier. For 1D (3D) systems, however, the conductivity-diffusion ratio is inversely (directly) proportional to the square root of the thermal energy $(k_BT)^{-(1/2)}$ ($(k_BT)^{(1/2)}$).

From the general charge density and diffusion-mobility expressions (eq. 4 and eq. 5, respectively), we can obtain the general conductivity relation for the semiconducting materials as,

$$\sigma_d \cong 2q^2 D \left(\frac{1}{\hbar}\right)^d \left(\frac{2m}{\pi k_B T}\right)^{1-(d/2)} \sum_{k=1}^{n}(-1)^{k+1} k^{1-(d/2)} \exp\left(\frac{k\eta}{k_B T}\right). \quad (14)$$

For high density systems, it is approximated to a simpler form as

$$\sigma_d \approx q^2 D \left(\frac{m}{2\pi\hbar^2}\right)^{d/2} (k_B T)^{(d/2)-1}. \quad (15)$$

As seen from eqs. (13), (14) and (15), the conductivity is directly related with the expressions for DOS, $D(E)|_d \propto E^{(d-2)/2}$, which is in agreement with well-established theorems. [2] Using this unified diffusive transport, one can relate the thermo-electric property with the help of WF law. Here, again the deterministic parameter is chemical potential. In this scenario, we have developed the expression of thermal conductivity ($\kappa$)-charge diffusion (*D*) ratio for degenerate 2D semiconductors, and it can be written as,

$$\frac{\kappa}{D} \approx \left(\frac{2\pi k_B}{3}\right)\frac{1}{\lambda^2}. \quad (16)$$

The presence of deep trap sites in the semiconductors introduces large energy deviation and in such cases the charge density can be calculated from Maxwell-Boltzmann distribution. This Maxwellian form also holds good for disordered or any non-degenerate semiconductors. In this context, our generalized diffusion-mobility ratio obeys the classical Einstein relation for all cases

$$\frac{D}{\mu} = \frac{k_B T}{q}. \tag{17}$$

The detail of derivation of diffusion-mobility ratio for relativistic cases goes as follows. The generalized charge density of the relativistic domain ($n_{d,rel}$) can be expressed as,

$$n_{d=1,2,rel} = \left(\frac{2}{\pi}\right)\left(\frac{k_B T}{\hbar c}\right)^d \sum_{k=1}^{n}(-1)^{k+1}\left(\frac{1}{k}\right)^d \exp\left(\frac{k\eta}{k_B T}\right) \tag{18}$$

where, c is the velocity of the light. The above density relation (eq. 18) is applicable for 1D and 2D materials. For 3D systems, the density can be written as

$$n_{d=3,rel} = \left(\frac{4}{\pi^2}\right)\left(\frac{k_B T}{\hbar c}\right)^3 \sum_{k=1}^{n}(-1)^{k+1}\left(\frac{1}{k}\right)^3 \exp\left(\frac{k\eta}{k_B T}\right) \tag{19}$$

Substituting eq. (18) and (19) in eq. (2), one can obtain the generalized diffusion-mobility relation in relativistic domain as,

$$\left.\frac{D}{\mu}\right|_{d,rel} = \frac{k_B T}{q}\left[\frac{\sum_{k=1}^{n}(-1)^{k+1}(k)^{-d}\exp\left(\frac{k\eta}{k_B T}\right)}{\sum_{k=1}^{n}(-1)^{k+1}(k)^{1-d}\exp\left(\frac{k\eta}{k_B T}\right)}\right]. \tag{20}$$

for d=1,2 and 3. In the case of relativistic 1D materials, we can formulate the charge density expression directly without invoking any approximation and can be written as,

$$n_{d=1,rel} = \frac{2k_B T}{\pi \hbar c}\ln\left(1+\exp\left(\frac{\eta}{k_B T}\right)\right), \tag{21}$$

Substituting eq. (21) in eq. (2), one can obtain the diffusion-mobility relation for 1D in the relativistic limit as,

$$\left.\frac{D}{\mu}\right|_{d=1,rel} = \frac{k_B T}{q}\left[\frac{\left(1+\exp\left(\frac{\eta}{k_B T}\right)\right)\ln\left(1+\exp\left(\frac{\eta}{k_B T}\right)\right)}{\exp\left(\frac{\eta}{k_B T}\right)}\right] \tag{22}$$

Applying Stirling approximation in case of 2D and 3D cases, one can derive the diffusion-mobility ratio as,

$$\left.\frac{D}{\mu}\right|_{d,rel} \approx \frac{k_B T}{q}\left[1 + \exp\left(\frac{\eta_{d+1}}{k_B T}\right)\right]. \qquad (23)$$

The above equation is the best approximated generalized D/□ formula for 1D, 2D and 3D systems in the relativistic limit. Here, $\eta_{d+1}$ is the general dimensional dependent chemical potential in the relativistic domain. In case of vanishing charge density limit and for nondegenerate systems, the chemical potential $\eta_{d+1} \to -\infty$, [28] and hence the above unified diffusion-mobility formula can be transformed to the classical Einstein relation as

$$\left.\frac{D}{\mu}\right|_{d,rel} \approx \frac{k_B T}{q}. \qquad (24)$$

It is to be noted that the relativistic form of charge density and diffusion-mobility equations for one-dimensional electrical transport (see eqs. (21) and (22)) is similar to nonrelativistic form for 2D semiconductors (see eqs (6) and (7)). In relativistic condition, transport in 1D materials has 2 (or 1+1) dimensional effect due to the existence of space-time which additionally contributes one more dimension. This additional dimension plays a central role for fast carrier dynamics in materials like, graphene. The same scenario can be extended to the relativistic carrier motion in 2D and 3D materials.

The variation of energy gap due to disorder can be calculated using disorder dependent energy dispersion equation and it can be used in Fermi-Dirac (F-D) distribution functions to address the disordered transport. [34] The product of two functions (using F-D) $g(E) = f(E)[1 - f(E)]$ provides the Gaussian like charge distribution. In fact, for disorder semiconductors, the product function $g(E)_S = f(E)_S[1 - f(E)_S]$ can deviate from the original Gaussian shape, leading to dispersive transport, which is in well-agreement with the earlier findings. [9, 39] The main point is that our unified theory can explain all the earlier experimental works on quantum transport, at both the relativistic and nonrelativistic limit, and at all temperature range in all three dimensions. It also reproduces Einstein equation in classical regime.

In summary, we have derived unified quantum classical relation of diffusion-mobility ratio for 1D, 2D and 3D materials with wide range of temperatures. The derived expressions dictate the nonlinear transport phenomena with respect to temperature in degenerate systems, and it can be reduced to the linear like transport with temperature in nondegenerate cases. Interestingly, all the quantities, like the electronic information of the systems, coupled strength of electric and thermal energy quantities like charge-heat current, are tunable by one parameter, i.e. chemical potential. Our generalized expressions fit quite well in the classical and quantum domains in both the nonrelativistic and relativistic cases. In reality, there exists degenerate materials due to the environmental interactions and such environmental variables can be accounted for by the chemical potential parameter which exactly satisfies the degenerate (or disorder) cases of nonequilibrium charge transport. We are able to relate the energy transport during the charge transport phenomena in degenerate cases which holds good corroboration with predictions by Tessler *et al*. [3, 4] Using our unified theory, one can explain the validity as well as the limitations of both Wiedemann-Franz law and Mott relation. Importantly, this letter provides the generalized version of diffusion-mobility ratio for low density as well as high density materials in both equilibrium and non-equilibrium situations.